\documentclass[showpacs,twocolumn,prl]{revtex4}

\usepackage{amsfonts}
\usepackage{amssymb}
\usepackage{amsmath}
\usepackage{graphicx}
\usepackage{color}
\usepackage{epsfig}

\begin{document}

\title{Long range failure-tolerant entanglement distribution}

\begin{abstract}
We introduce a protocol to distribute entanglement between remote parties.
Our protocol is based on a chain of repeater stations, and exploits topological encoding to tolerate very high levels of defects and errors.
The repeater stations may employ probabilistic entanglement operations which usually fail; ours is the first protocol to explicitly allow for technologies of this kind.
Given an error rate between stations in excess of $10\%$, arbitrarily long range high fidelity entanglement distribution is possible even if the heralded failure rate within the stations is as high as $99\%$, providing that unheralded errors are low (order $0.01\%$).
\end{abstract}

\author{Ying Li}
\affiliation{Centre for Quantum Technologies, National University of Singapore, 3 Science Drive 2, Singapore 117543}
\author{Sean D. Barrett}
\affiliation{Blackett Laboratory and Institute for Mathematical Sciences, Imperial College London, London SW7 2PG, United Kingdom}
\author{Thomas M. Stace}
\affiliation{School of Mathematics and Physics, University of
Queensland, Brisbane, QLD 4072, Australia}
\affiliation{Centre for Quantum Technologies, National University of Singapore, 3 Science Drive 2, Singapore 117543}
\author{Simon C. Benjamin}
\affiliation{Department of Materials, University of Oxford, Parks Road, Oxford OX1 3PH, UK}
\affiliation{Centre for Quantum Technologies, National University of Singapore, 3 Science Drive 2, Singapore 117543}

\maketitle

\textit{Introduction.}
Distributing an entangled state among remote quantum computers is one of the fundamental tasks of quantum information technologies.
It is crucial for quantum teleportation, quantum cryptography and distributed quantum computing.
Using direct transmission, the success probability of transmitting a qubit and the fidelity of the resulting quantum state decrease exponentially with distance.
Therefore, one needs quantum repeaters to achieve long distance entanglement \cite{repeater,DLCZ}.
A good quantum repeater protocol should be fault-tolerant and support a high communication rate.
In this paper, we will propose a protocol to distribute entanglement between two remote quantum computers.
We consider noise in quantum communication channels, and of course errors generated by operations within the repeaters.
We assume that the repeater stations may employ non-deterministic entanglement operations (EOs): that is, a means of entanglement, even within the a single repeater, that often fails but the failures are `heralded'.
In addition there is of course a finite error rate even for the operations that are deemed successful.
Non-deterministic EOs will occur within individual repeater stations if, for example, their internal hardware is based on networking small quantum registers together optically, i.e. qubits can be entangled by joint measurements on single photons emitted from these qubits rather than control of interactions \cite{NVC,ion}.
Such an architecture may be much easier to implement in a scalable way than monolithic architectures e.g. large scale ion traps.
Even with this assumption that EOs fail both between {\em and within} repeater stations, we find that the rate of distributing entanglement decreases only logarithmically with the communication distance.

\begin{figure}[tbp]
\includegraphics[width=8 cm]{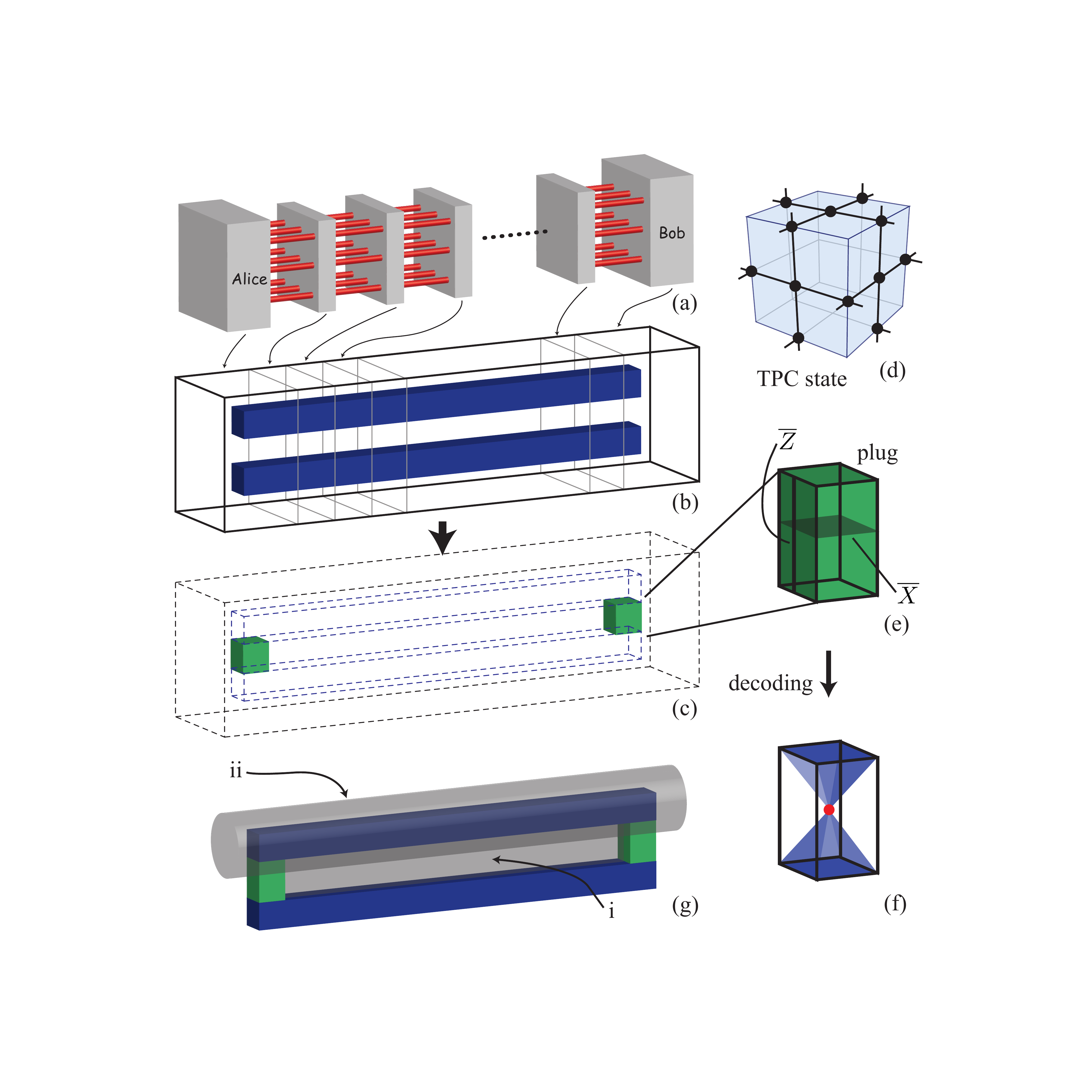}
\caption{
The scheme of quantum entanglement distribution protocol based on topologically protected cluster (TPC) state.
(a) Alice and Bob can be entangled via a chain of quantum repeater stations, which are connected by optical quantum communication channels.
(b) Each station contains a `slice' of the TPC state.
The TPC state contains two empty tubes (blue) without any qubit.
(c) Once the TPC is complete, all qubits are measured in $X$ basis except two parts of the TPC state (green) in stations of Alice and Bob respectively; these are called plugs and contain the eventual encoded shared Bell pair.
(d) The elementary cell of the TPC state.
Each logical qubit is encoded as subfigure (e) and can be decoded as subfigure (f) (see text).
(g) Two surfaces propagating correlations between two logical qubits.
}
\label{scheme}
\end{figure}

{Cluster states are resources of measurement-based quantum computing \cite{MBQC}, and long-range entanglement can be established in noisy cluster states \cite{longrange}.
In this paper, we propose a protocol of distributing entanglement by single-qubit measurements on a topologically protected cluster (TPC) state \cite{R.Raussendorf} across the chain of repeater stations.}
The TPC state must first be grown via operations within repeaters together with quantum communication between pairs of neighboring repeaters.
The operations within repeaters are expected to have a much better performance than communications between repeaters (since the latter may be over distances of kilometres).
We find that the protocol is valid if the probability of an error occurring in the communication channel is lower than a threshold, which is $15\%$ when errors induced by operations within repeaters are negligible.
With errors less than the threshold, entanglement can be established between two remote logical qubits encoded in two separated graph states, which may be used for further information processing via the topological measurement based quantum computing \cite{R.Raussendorf}.
Alternatively one can also decode each logical qubit to a physical qubit via single-qubit measurements.
Although we describe only the two-party protocol here, it should be straightforward to generalize for distributing multi-party entanglement.

In this protocol, the quality of the eventual entanglement between logical qubits is only limited by the number of qubits in each repeater.
Therefore, our protocol effectively distills as well as distributes entanglement.
The idea of using an error correction code with protected logical qubits for remote entanglement was firstly reported in Ref. \cite{Lukin}, in which the Calderbank-Shor-Steane code is employed. Subsequently  3D lattice-based distribution has also been studied~\cite{Perseguers} and the extension to lower dimensionality has been examined~\cite{Grudka}.
Recently, in a protocol for quantum state transfer of a surface-code-encoded qubit, the efficiency of quantum communication is greatly improved by removing the necessity of two-way communication \cite{surface}.
Compared with these protocols, ours is the first to consider a probabilistic architecture within each repeater station, so that the entanglement distribution can be efficient even if EOs are far from deterministic.

\textit{Quantum Repeaters based on Cluster States.}
Alice and Bob are entangled via a chain of quantum repeater stations.
Two nearby repeaters are connected by optical quantum communication channels [Fig. \ref{scheme}(a)] --  essentially a bundle of optical fibres that are used in parallel. 
To give an overview of the process:
Firstly, a TPC state is grown across quantum repeater stations via probabilistic EOs and quantum communications between nearby stations.
The TPC state contains two parallel empty tubes, which terminate in stations of Alice and Bob.
Each empty tube is a void in the TPC state, with an elongated shape and shown as a blue rectangular cuboid in Fig. \ref{scheme}(b).
Once the TPC state is generated, measurements in the $X$ basis are performed on all qubits except two parts of the TPC state located in stations of Alice and Bob respectively [see Fig. \ref{scheme}(c)].
The two parts which are to remain unmeasured are called plugs, and are connected with empty tubes.
Two empty tubes and two plugs form a closed loop.
There is one logical qubit encoded in each plug.
After all other quits are measured, and the outcomes are communicated to Alice and Bob, then these two logical qubits are entangled as one of the Bell states (determined by measurement outcomes).

The TPC state is a cluster state of qubits located on the a cubic lattice \cite{R.Raussendorf}.
There is one qubit on each face and edge of the elementary cell [Fig. \ref{scheme}(d)].
By shifting the lattice, one can transfer qubits on faces to edges, and vice versa.
The new lattice is called the dual lattice of the original primal lattice.
The TPC state is stabilized by $K(c)=\prod_{a\in c}X_{a}\prod_{b\in \partial c}Z_{b}$, where $c$ is an arbitrary primal (dual) surface and $\partial c$ is the primal (dual) chain as the boundary of $c$.
Qubits in the set $c$ ($\partial c$) are located on faces (edges) composing the surface (chain) $c$ ($\partial c$).
The logical qubit is encoded in a plug as $\overline{X}=\prod_{a\in section}X_{a}$ and $\overline{Z}=\prod_{b\in line}Z_{b}$, where $\overline{X}$ and $\overline{Z}$ are Pauli operators of the logical qubit.
Here, \textit{section} is a dual surface across the plug, and \textit{line} is a primal chain on the surface of the plug and connecting two empty tubes [Fig. \ref{scheme}(e)].
We consider two stabilizers according to the following surfaces: (i) $c_{\mathrm{i}}$ is a primal surface whose boundary is enclosed by the tube-plug loop, and (ii) $c_{\mathrm{ii}}$ is a closed dual surface enveloping one empty tube and crossing two plugs [Fig. \ref{scheme}(g)].
The two stabilizers are $K(c_{\mathrm{i}})=\overline{Z}_{A}\overline{Z}_{B}\prod_{a\in c_{\mathrm{i}}}X_{a}$ and $K(c_{\mathrm{ii}})=\overline{X}_{A}\overline{X}_{B}\prod_{a\in c_{\mathrm{ii}}^{\prime }}X_{a}$, where $A,B$ denote Alice and Bob respectively, and $c_{\mathrm{ii}}^{\prime }$ denotes the part of the surface $c_{\mathrm{ii}}$ outside two plugs.
After measurements in the $X$ basis, one can replace $X_{a}$ with measurement outcomes.
Then, we get two new stabilizers $\overline{Z}_{A}\overline{Z}_{B}=\pm 1$ and $\overline{X}_{A}\overline{X}_{B}=\pm 1$, i.e. the two logical qubits are stabilized as one of Bell states.
Here, the two signs depend on measurement outcomes.

{Besides two-party entanglement, we note that our scheme can be directly generalized to multi-party entanglement, e.g. three-party and four-party entanglement as shown in Ref. \cite{SupMat}.}

Noise in quantum communication channels and imperfections in operations will give rise to phase errors on the TPC state.
In order to eliminate errors from the Bell state of two logical qubits, we monitor errors on the TPC state by parity check operators $K(c_{c})$, where $c_{c}$ are minimum closed surfaces.
Usually, minimum closed surfaces are surfaces of elementary cubes.
However, some qubits on the TPC state may be missing.
The parity check operator of an elementary cube with missing qubits can not be used to detect errors.
Then, one has to use products of parity check operators connected by missing qubits to form a new set of parity check operators \cite{Sean}.
Parity check operators reveal the endpoints of error chains, where an error chain (ring) is a sequence of phase errors.
If the number of phase errors on the surface $c_{c}$ is odd, the existence of errors can be identified by $K(c_{c})$, which is called an error syndrome.
Errors are not actively corrected, rather parities of $\prod_{a\in c_{\mathrm{i}}}X_{a}$ and $\prod_{a\in c_{\mathrm{ii}}^{\prime }}X_{a}$, are modified by knowledge of the total number of error chains crossing surfaces $c_{\mathrm{i}}$ and $c_{\mathrm{ii}}^{\prime }$ respectively.
After the error correction, only error rings encircling  the tube-plug loop, error chains connecting two empty tubes and error chains connecting the loop with the boundary of the TPC state \cite{SupMat}, may contribute an error on logical qubits.
If noise and imperfections are less than a threshold, the probability of an error on logical qubits decreases exponentially with the minimum length of these error rings and error chains \cite{R.Raussendorf}.

\begin{figure}[tbp]
\includegraphics[width=8.6 cm]{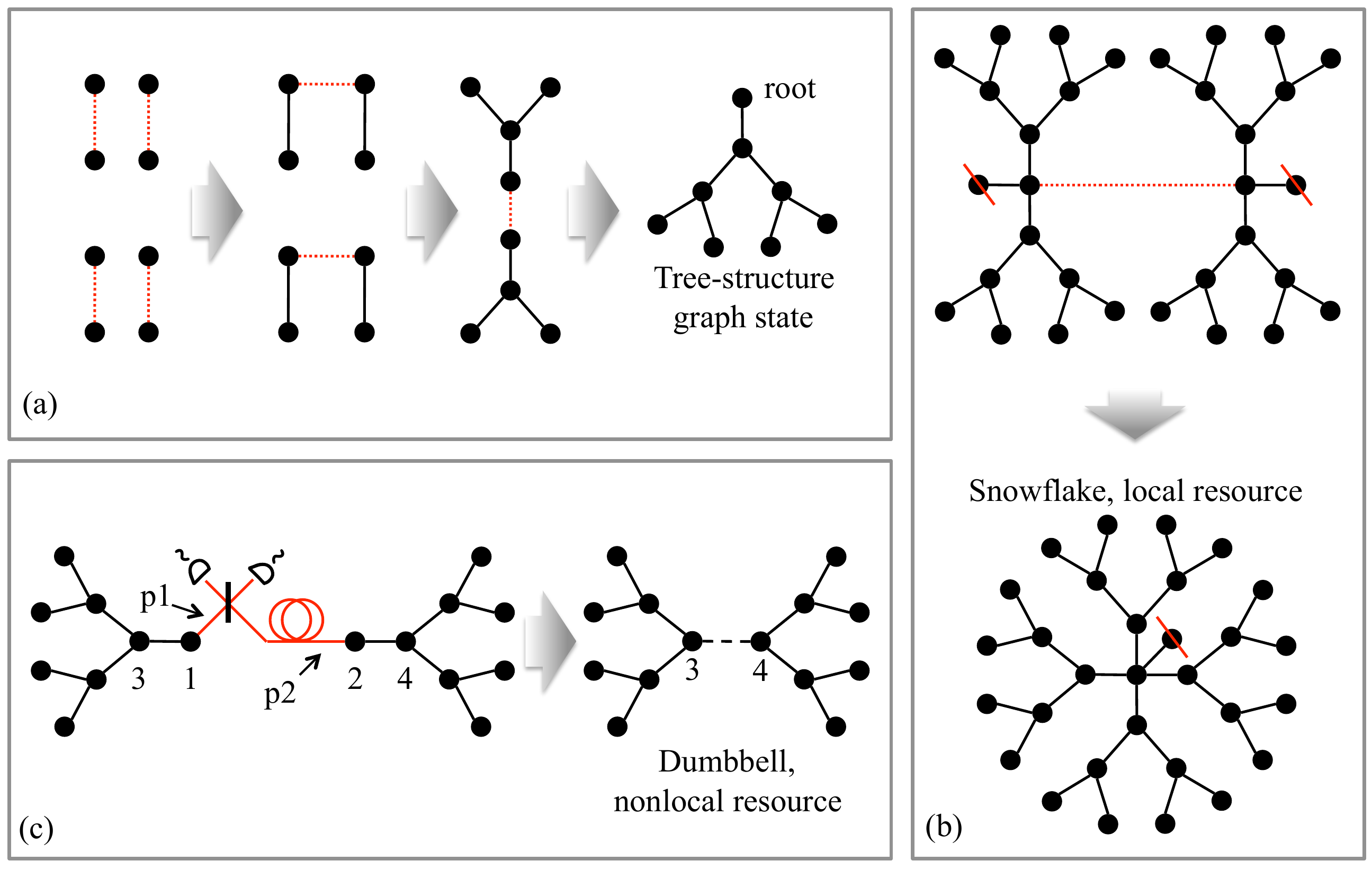}
\caption{
Resource graph states, i.e. building-blocks, for growing the topologically protected cluster state.
Each red dotted line denotes a parity projection (PP).
(a) Tree graph states can be grown by PPs on roots of trees.
(b) Four trees can be fused into a `snowflake' graph state as the following: fusing each pair of trees into a bigger tree at first; cutting two roots by measurements in $Z$ basis; fusing them into a snowflake and cutting the unwanted qubit.
(c) Two trees in different quantum repeater stations are fused into a {\em dumbbell} graph state by a Bell measurement on photon-$\mathrm{p}1$ and the photon-$\mathrm{p}2$, each associated with a qubit in a different stations.
One of the photons ($\mathrm{p}2$) will have travelled between stations.
The Bell measurement is followed by a measurement in the $Y$ basis on the qubit-$1$ and a measurement in the $X$ basis on the qubit-$2$, in order to get the desired dumbbell graph state.
}
\label{structure}
\end{figure}

\textit{Cluster State Growth.-}
In order to grow the TPC state across quantum repeater stations, some `building-block' graph states should first be prepared within each repeater device. It is through the use of these building-blocks that we overcome the impact of high EO failure rates when we create the large scale TPC state.
The structure of these elementary graph states can be a star \cite{Nielsen}, a line \cite{line}, a cross \cite{DuanRR}, or a tree \cite{Bodiya,UT}.
In this paper, we take the tree structure as an example, and the scheme can be adapted to other structures.
The tree structure accumulates fewer errors than other structures when the success probability of EOs is low \cite{UT,YingLi2010}.
Tree-structure graph states can be generated by using parity projections (PPs) \cite{NVC}.
A PP on roots of two individual trees can fuse them into a double-size tree [Fig. \ref{structure}(a)].
If all PPs are successful, after $n$ steps, one can grow a tree with $2^{n}$ qubits from separated qubits, where the integer $n$ is called the generation of the tree.

Trees are fused into two kinds of building-block graph states.
\textit{Snowflake} graph states are prepared by fusing four trees [Fig. \ref{structure}(b)].
Each snowflake will ultimately correspond to a specific qubit on the TPC state.
Each quarter of a snowflake is used to establish a connection with a neighboring snowflake.
We refer to the second kind of building-block as a \textit{dumbbell}.
These are nonlocal building blocks connecting two nearby quantum repeater stations [Fig. \ref{structure}(c)].
A dumbbell is formed by two trees located in different stations.
For example, suppose that the basic qubits are optically active atoms: then in order to prepare a dumbbell, we cause each root qubit emit a single photon as $\left\vert \eta \right\rangle_{j}\rightarrow \left\vert \eta \right\rangle _{j}\left\vert \eta
\right\rangle _{\mathrm{p}j}$, where $j=1,2$ denotes a root qubit, `$\mathrm{p%
}j$' denotes the corresponding photonic qubit, $\eta =0,1$ is the label the state in the computational basis and the photonic qubit can be encoded in polarization, frequency \cite{trappedatoms}\ or time-bin \cite{SDBPK}.
One photon is transmitted from one station to another.
After a Bell measurement on two photons and single-qubit measurements on roots, we obtain the dumbbell graph state \cite{SupMat}.

Making a building-block graph state requires all operations to be successful, whose probability may be quite small.
Therefore, building-block graph states are produced with a post selection strategy: if an operation is heralded as failed, the corresponding graph state is abandoned with the qubits reinitialized.

\begin{figure}[tbp]
\includegraphics[width=7 cm]{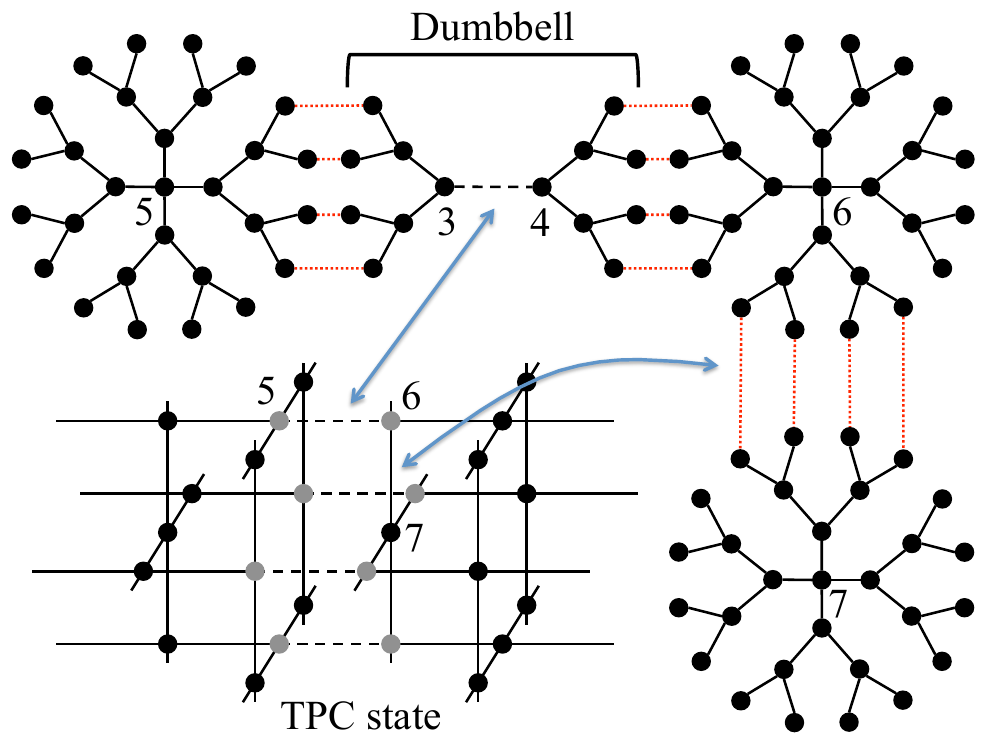}
\caption{
The strategy of assembling resource graph states into the full topologically protected cluster (TPC) state which spans all  quantum repeater stations.
(a) Snowflakes within the same station are connected directly to each other by parity projections (red dotted lines) on leaves.
Two snowflakes in different stations can be connected via a dumbbell which incorporates the required nonlocal connection (dash line).
(b) After extraneous quits are removed, ultimatelty the qubits at the heart of each snowflake survive as nodes of the TPC state.
}
\label{construction}
\end{figure}

Once a sufficient number of each resource (snowflakes and dumbbells) have been generated, we can assemble them to create a suitable TPC state. Snowflakes are assembled by PPs on leaves, which are qubits on the edge of a snowflake (Fig. \ref{construction}).
Two snowflakes in the same quantum repeater station can be connected directly, while two snowflakes in different stations are connected by bridging them with a dumbbell shared by these two stations.
The number of leaf qubits on each quarter of a snowflake is $2^{n-1}$.
Therefore, the failure probability of connecting two snowflakes in the same station is $F_{\mathrm{L}}=f^{2^{n-1}}$, and the failure probability of connecting two snowflakes in different stations is $F_{\mathrm{NL}}\simeq 2F_{\mathrm{L}}$, where $f$ is the basic failure probability of EOs.
After establishing connections between snowflakes, all qubits except those at the center of each snowflakes are removed by appropriate single-qubit measurements, so that the surviving qubits form the TPC state.
Here, the measurement pattern for removing qubits can be found in Ref. \cite{SupMat}.
Since some snowflakes have failed to connect, this implies some missing connections on the TPC state.
We presently describe simulations establishing that when connections are rarely missing, i.e. $F_{\mathrm{L}}<5\%$, then the cluster state is {\em well connected}: it is easy to find surfaces propagating correlations between two logical qubits, indeed this is guaranteed in the scaling limit (as expected from percolation theory) \cite{percolation,Sean}.

As a footnote to this section we note that the `building-block' strategy is not always necessary.
If the failure probability of EOs is low enough $f<5\%$, one may generate the TPC state directly, for example, by using control phase gates \cite{R.Raussendorf}, where control-phase gates on two qubits located in different quantum repeater stations can be simulated by consuming entanglement prepared via quantum communication \cite{Plenio}. However here we are interested in the general case where the failure probability may be very high. 

\textit{Noise, Imperfections and Error Correction.-}
Both noise in quantum communication channels and imperfections in operations can give rise to errors on the TPC state.
We assume communication noise is depolarized, and described by the superoperator $E=(1-\epsilon )[1_{\mathrm{p}2}]+\epsilon ([X_{\mathrm{p}2}]+[Y_{\mathrm{p}2}]+[Z_{\mathrm{p}2}])/3$ [see Fig. \ref{structure}(c)].
We call qubits with nonlocal connections  `joint qubits' [gray circles in Fig. \ref{construction}].
Errors induced by communication noise may make phase errors on corresponding joint qubits (qubits $5$ and $6$) with a probability $2\epsilon /3$ for each of them \cite{SupMat}.
Consider first the case that internal operations within stations are perfect (when heralded as successful); then only joint qubits have errors, and these imperfect qubits exist in specific non-adjacent layers of the TPC state. Then error correction can be performed independently on each such layer.
The error threshold of a two dimensional layer is about $10\%$ in the limit of a perfectly connected lattice \cite{threshold2D}.
Moreover a near-perfectly connected lattice would indeed be achievable since, given error free EOs within repeaters, one could always grow sufficiently big tree structures to make $F_{\mathrm{L}}$ as low as desired.
Therefore, with perfect operations, the condition of getting a correct correlation between two logical qubits faithfully is $2\epsilon /3\lesssim 10\%$, i.e. the error threshold of communication noise is $\epsilon_{\mathrm{t}}\simeq15\%$.

\begin{figure}[tbp]
\includegraphics[width=8.5 cm]{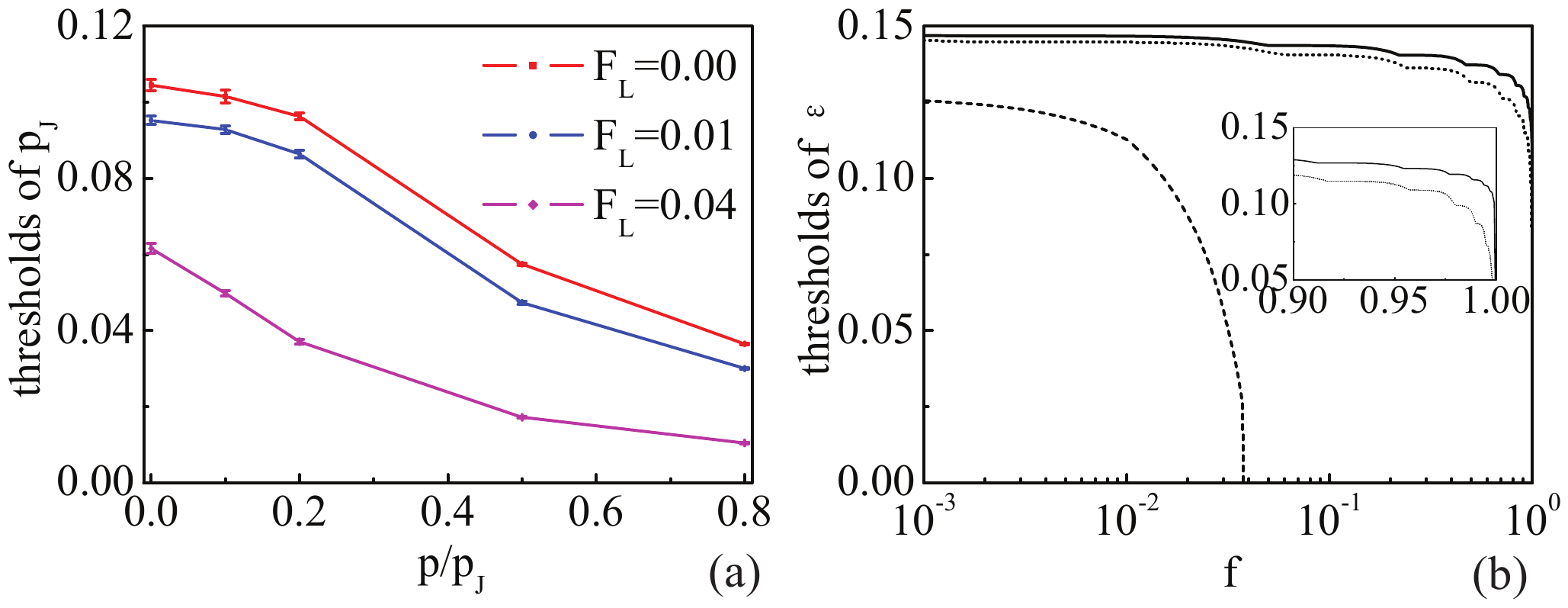}
\caption{
Thresholds of error correction on the topologically protected cluster (TPC) state.
(a) Thresholds of phase errors on joint qubits, which is dependent on the ratio between the error probability on joint qubits ($p_{J}$) and the error probability on other qubits ($p$).
(b) Thresholds of communication noise $\epsilon$ with operational error rate $10^{-4}$ (solid line), evaluated from the linear interpolation of data in subfigure (a).
By using control-phase gates to generate the TPC state directly, the error rate can be much higher ($2\times 10^{-3}$) but only a failure probability ($f$) lower than $4\%$ is tolerable (dash line).
Here we have assumed that memory errors happen at a lower rate than operational errors.
Memory errors at 10\% of the operational error rate can lower the threshold, but not dramatically (dotted line).
}
\label{thresholds}
\end{figure}

With imperfect operations, all qubits on the TPC state may affected by phase errors.
If the distribution of phase errors is uniform, i.e. all qubits may have a phase error with the same probability, the  threshold of phase errors is about $3\%$ for perfectly connected TPC state \cite{threshold}.
However, in our case, the TPC state grown by probabilistic EOs is unlikely to be perfectly connected and there are more errors on joint qubits than others.
Our strategy is to treat missing connections by transforming them to qubit loss, by means of deleting the qubits with missing connections using measurements in the $Z$ basis.
Then, the loss probability of joint qubits is $5F_{\mathrm{L}}$, and the loss probability of other qubits is $4F_{\mathrm{L}}$.
We determine error thresholds for general cases numerically as shown in Fig. \ref{thresholds}(a), using the method developed in Ref. \cite{Tom,Sean}.

The error rate of imperfect operations must be lower than the threshold of fault-tolerant quantum computing (FTQC).
The threshold of FTQC on the TPC state with non-deterministic EOs (deterministic control-phase gates) is about $2\times 10^{-4}$ \cite{YingLi2010} ($5\times 10^{-3}$ \cite{R.Raussendorf}).
By optimizing the size of trees, (a bigger tree can reduce missing connections but generate more errors), we have obtained the thresholds of tolerable communication noise in the presence of finite error rates for internal EOs, see Fig. \ref{thresholds}(b).
If the error rate of operations is $10^{-4}$, the threshold of communication noise is about $11\%$ when the success probability of entangling operations is $1\%$.
In contrast, by using control-phase gates to generate the TPC state directly, the threshold of communication noise is still above $10\%$ even if the error rate of operation is $2\times 10^{-3}$, but the success probability must be higher than $98\%$.


\textit{Full decoding.-}
A logical qubit can be decoded into a physical qubit by measurements on the corresponding plug, leaving just one qubit unmeasured.
The residual qubit carries the quantum state of the logical qubit.
For decoding, two (blue) pyramids inside the plug, whose apexes hold the residual qubit (red circle) and bases connect tubes, are measured in the $Z$ basis, while other qubits are measured in the $X$ basis [see Fig. \ref{scheme}(f)].
The residual qubit can acquire an error if there is an error chain connecting two pyramids.
Therefore, the probability of an error on the residual qubit is $p+O(p^{3})$ \cite{R.Raussendorf}, where $p$ is the probability of phase errors on the residual qubit, which is usually lower than $3\%$.

\textit{Performance.-}
The probability of errors on two entangled logical qubits decreases exponentially with the minimum length of error rings and error chains \cite{R.Raussendorf}.
We design the TPC state as follows: the perimeters of two empty tubes, the distance between empty tubes, and the distance between each empty tube and the boundary, are each proportional to the same length scale $L$.
The length of the TPC state, i.e., the number of quantum repeater stations, can increase the probability of error rings and error chains linearly \cite{longrange}.
Therefore, the overall probability of errors on two entangled logical qubits is $\epsilon _{\mathrm{E}}\propto Ne^{-\kappa L}$, where $N$ is the number of stations, $\kappa $ is a constant depending on $p$, $p_{J}$ and $F_{\mathrm{L}}$.
To achieve a given quality of entanglement, we need a TPCS with $L=O(\log (N/\epsilon _{\mathrm{E}})/\kappa )$.
The number of photonic qubits transferred between two nearby stations is proportional to $L^{2}$.
Therefore the overall entanglement distribution rate of our scheme is $R_{N}=O(\log ^{-2}(N/\epsilon _{\mathrm{E}})/\kappa )$.

\bigskip

In conclusion, we have described an advanced protocol for distributing entanglement through the use of repeater stations which together generate a topologically protected cluster state.
We find that the approach is remarkably robust to errors, while the resource cost within each repeater scales only logarithmically with the total distance over which entanglement is to be shared.

While preparing this document we became aware of a manuscript  describing closely related research:
Ashley Stephens, Jingjing Huang, Kae Nemoto and William J. Munro, ``Fault-tolerant quantum communication with rare-earth elements and superconducting circuits".


\begin{thebibliography}{99}
\bibitem{repeater} H.-J. Briegel \textit{et al.}, Phys. Rev. Lett. \textbf{81}, 5932 (1998).

\bibitem{DLCZ} L.-M. Duan \textit{et al.}, Nature \textbf{414}, 413 (2001).

\bibitem{NVC} S. C. Benjamin, B. W. Lovett, and J. M. Smith, Laser \& Photonics Reviews, \textbf{3}, 556 (2009).

\bibitem{ion} D. L. Moehring \textit{et al.}, J. Opt. Soc. Am. B \textbf{24}, 300 (2007).

{
\bibitem{MBQC} R. Raussendorf and H. J. Briegel, Phys. Rev. Lett. \textbf{86}, 5188 (2001);
R. Raussendorf, D. E. Browne, and H. J. Briegel, Phys. Rev. A \textbf{68}, 022312 (2003).

\bibitem{longrange} Robert Raussendorf, Sergey Bravyi, and Jim Harrington, Phys. Rev. A \textbf{71}, 062313 (2005).
}

\bibitem{R.Raussendorf} R. Raussendorf, J. Harrington, and K. Goyal, Ann. Phys. \textbf{321}, 2242 (2006);
R. Raussendorf and J. Harrington, Phys. Rev. Lett. \textbf{98}, 190504 (2007);
R. Raussendorf, J. Harrington, and K. Goyal, N. J. Phys. \textbf{9}, 199 (2007).

\bibitem{Lukin} Liang Jiang \textit{et al.}, Phys. Rev. A. \textbf{79}, 032325 (2009).

\bibitem{Perseguers}S. Perseguers, Phys. Rev. A {\bf 81}, 012310 (2010).

\bibitem{Grudka} A. Grudka {\em et al}  arXiv:1202.1016 [quant-ph].

\bibitem{surface} A. G. Fowler \textit{et al.}, Phys. Rev. Lett. {\bf 104}, 180503 (2010).

\bibitem{SupMat} Supplementary material, http://qunat.org/papers/topCom

\bibitem{Nielsen} M. Nielsen, Phys. Rev. Lett. \textbf{95}, 080503 (2005).

\bibitem{line} S. D. Barrett and P. Kok, Phys. Rev. A \textbf{71}, 060310 (2005);
S. C. Benjamin, Phys. Rev. A \textbf{72}, 056302 (2005).

\bibitem{DuanRR} L.-M. Duan and R. Raussendorf, Phys. Rev. Lett. \textbf{95}, 080503 (2005).

\bibitem{Bodiya} T. P. Bodiya and L.-M. Duan, Phys. Rev. Lett. \textbf{97}, 143601 (2006).

\bibitem{UT} Y. Matsuzaki, S. C. Benjamin, and J. Fitzsimons, Phys. Rev. Lett. \textbf{104}, 050501 (2010).

\bibitem{YingLi2010} Y. Li, S. D. Barrett, T. M. Stace, and S. C. Benjamin, Phys. Rev. Lett. \textbf{105}, 250502 (2010).

\bibitem{trappedatoms} D. L. Moehring, M. J. Madsen, K. C. Younge, R. N. Kohn, Jr., P. Maunz, L.-M. Duan, and C. Monroe, J. Opt. Soc. Am. B \textbf{24}, 300 (2007).

\bibitem{SDBPK} S. D. Barrett and P. Kok, Phys. Rev. A \textbf{71}, 060310 (2005).

\bibitem{percolation} C. D. Lorenz and R. M. Zi , Phys. Rev. E \textbf{57}, 230 (1998).

\bibitem{Tom} T. M. Stace, S. D. Barrett, A. C. Doherty, Phys. Rev. Lett. \textbf{102}, 200501 (2009);
T. M. Stace, S. D. Barrett, Phys. Rev. A \textbf{81}, 022317 (2010).

\bibitem{Sean} S. D. Barrett, T. M. Stace, Phys. Rev. Lett. \textbf{105}, 200502 (2010).

\bibitem{Plenio} J. Eisert, K. Jacobs, P. Papadopoulos, and M. B. Plenio, Phys. Rev. A \textbf{62}, 052317 (2000).

\bibitem{threshold} T. Ohno, G. Arakawa, I. Ichinose and T. Matsui, Nucl. Phys. B \textbf{697}, 462 (2004).

\bibitem{threshold2D} C. Wang, J. Harrington, and J. Preskill, Annals of Physics \textbf{303}, 31 (2003);
F. Merz and J. T. Chalker, Phys. Rev. B \textbf{65}, 054425 (2002).

\end{thebibliography}
\end{document}